\documentclass{ws-mplb}
\pdfoutput=1
\usepackage{url}
\usepackage{amsfonts,amsmath,amssymb}
\usepackage{graphicx}
\usepackage[square,numbers,sectionbib]{natbib}
\usepackage{chemformula}
\begin{document}

\markboth{Saurabh Kumar Srivastav and Anindya Das}
{Thermal conductance of QH state...}

%
\catchline{}{}{}{}{}
%

\title{Quantized heat flow in graphene quantum Hall phases: Probing the topological order}

\author{Saurabh Kumar Srivastav}

\address{Department of Condensed Matter Physics, Weizmann Institute of Science\\
Rehovot, Israel 76001,
Israel\\
ssrivastav210@weizmann.ac.il\\
https://orcid.org/0000-0002-0498-4217}

\author{Anindya Das}

\address{Department of Physics, Indian Institute of Science\\
Bangalore, 560012, India\\
anindya@iisc.ac.in\\
https://orcid.org/0000-0002-6310-1576}

\maketitle


\begin{abstract}
Topological quantum numbers are often used to
characterise the topological order of phase having protected gapless edge
modes when the system is kept in a space with the boundary. The famous examples in this category are the quantized electrical Hall conductance and thermal Hall conductance, which encodes the topological order of integer and fractional quantum Hall states. Here, we review the recent thermal transport study of integer and fractional quantum Hall states realized in graphene-based van der Waals heterostructures.  
\end{abstract}


\section{Introduction}
Topological properties of the electronic phases are often encoded in the quantized physical quantity, such as electrical Hall conductance and thermal Hall conductance. Although the notion of the topology in condensed matter physics was introduced long back in 1972 by J. Michael Kosterlitz and David J. Thouless\cite{kosterlitz1972long,kosterlitz2018ordering}, the first experimental discovery of the topological quantum liquid was realized by Klaus von Klitzing and coworkers in silicon field-effect transistor at low temperature and subjected to a high magnetic field. They found that the Hall conductance was precisely quantized in an integer multiple of the $G_0=\frac{e^2}{h}$\cite{klitzing1980new} (where h is Planck's constant and e is the elementary electronic charge.), accompanied by the vanishing longitudinal resistivity. This phenomenon is known as the integer quantum Hall (IQH) effect and is the first experimentally observed topological phase of electrons in a solid.  For this discovery,
Klaus von Klitzing received the Nobel Prize in Physics in 1985. With improved sample quality, in 1982, Tsui, Stormer, and Gossard observed that in addition to the integer quantum Hall plateau, a new plateau appeared beyond the conventional sequence of the integer quantum Hall state. It was quantized at the value $\rho_{xy} = \frac{h}{3e^2}$ accompanied by a minimum in $\rho_{xx}$\cite{tsui1982two}. Furthermore, with improvement in sample quality, the plateaus were observed at  $\rho_{xy} = \frac{h}{\nu e^2} = \frac{p}{q}\frac{h}{e^2}$ accompanied by vanishing $\rho_{xx}$, where $\nu$ is the filling factor, $p$ and $q$ are positive integer numbers. A few examples in this series are $\nu = 1/5, 1/3, 2/5, 3/7, 4/9, 5/9, 3/5, 4/7, 2/3,...$ in the lowest Landau level and $\nu = 4/3, 5/3, 7/5, 5/2, 12/5, 13/5,...$ in the higher Landau levels\cite{tsui1982two,tsui1999nobel,chang1984higher,clark1986odd,ebert1984fractional,goldman1988evidence,mendez1983high,mendez1984fractional,pan2003fractional,stormer1983fractional,willett1987observation,stormer1999fractional,eisenstein1990fractional}. The observation of the quantum Hall plateaus at the fractional filling is known as fractional quantum Hall (FQH) effect and lead to the another Nobel prize in physics in 1998. It is one of the most correlated states observed in condensed matter systems. In contrast to the integer quantum Hall states, a single-particle phenomenon, the fractional quantum Hall phase is a highly correlated state of the electrons. It emerges when the Landau levels are partially filled. 

The transition from one IQH or FQH plateau to another IQH or FQH plateau is a classic example of topological phase transition between two gapped bulk phases, each characterized by a different topological order. To probe the topological order of these phases, at first, one would naively expect to use an experimental technique that directly probes the bulk of these gapped phases. However, it turns out to be tedious and difficult to probe the bulk. Thanks to the bulk-edge correspondence principle, the topological order of bulk can also be accessed by studying the gapless edge modes at the physical boundary of the system, which is relatively easy to probe compared to the gapped bulk state. Theoretically, the gapless edge modes can be probed via quantized electrical Hall conductance and thermal Hall conductance. However, till now, most of our understanding of the topological order of IQH and FQH effects comes from electrical Hall conductance measurements, which is quite successful for the IQH phase and particle-like FQH phases, hosting only downstream edge modes. However, the transport properties become quite complex for hole-conjugate states hosting counter-propagating edge modes, which demands the measurement of quantized thermal Hall conductance. 

This article reviews the electronic thermal conductance study of the integer and several fractional quantum Hall states in graphene-based devices. By contrast to the conventional two-dimensional electron gas in \ch{GaAs/AlGaAs}, the linear, Dirac-like spectrum of low energy excitations and the pseudospin degeneracy make single-layer graphene (SLG) a unique, truly two-dimensional “relativistic” electronic system. The $\pi$ Berry's phase in graphene gives rise to Landau Levels structure, which is no longer evenly spaced. The Landau levels in graphene have four-fold internal degeneracy: two for spin and two for valley degree of freedom. As a result of the Dirac-like dynamics, the Hall conductivity $\sigma_{xy}$ exhibits an unconventional sequence of the quantization. Furthermore, the high mobility and gate tunability of the carrier density in graphene over a wide range provides a huge experimental phase space for the study of the symmetry-broken integer quantum Hall (IQH) and fractional quantum Hall phases (FQH). Since the unique band structure and the spin-valley degeneracy of graphene are quite different from the conventional \ch{GaAs/AlGaAs} 2D gas, it is worth starting with the basics of single-particle band structure, and resulting Landau Levels dispersion in the magnetic field. Next, we briefly describe the effect of the symmetry breaking and its signature in the transport experiments, mainly in electrical conductance. Then, we discuss the emergence of the fractional quantum Hall states in single-layer graphene. Next, we discuss the topological order of the QH states and their detection schemes. Later on, we provide a detailed discussion of the electronic thermal conductance measurement scheme of QH states, which is the central core of this review article. After the technical introduction of the measurement scheme, we will describe the results obtained from the three sets of thermal conductance measurements of integer and fractional quantum Hall states in single-layer and bilayer graphene. Finally, we end the review with a discussion on the impact of the experimental work discussed in this article and future endeavours. 

\section{Anomalous quantum Hall effect in graphene}
\subsection{Single particle band structure of `Single layer graphene}

\begin{figure}[t!]
 	\centering
 	\includegraphics[width=0.9\textwidth]{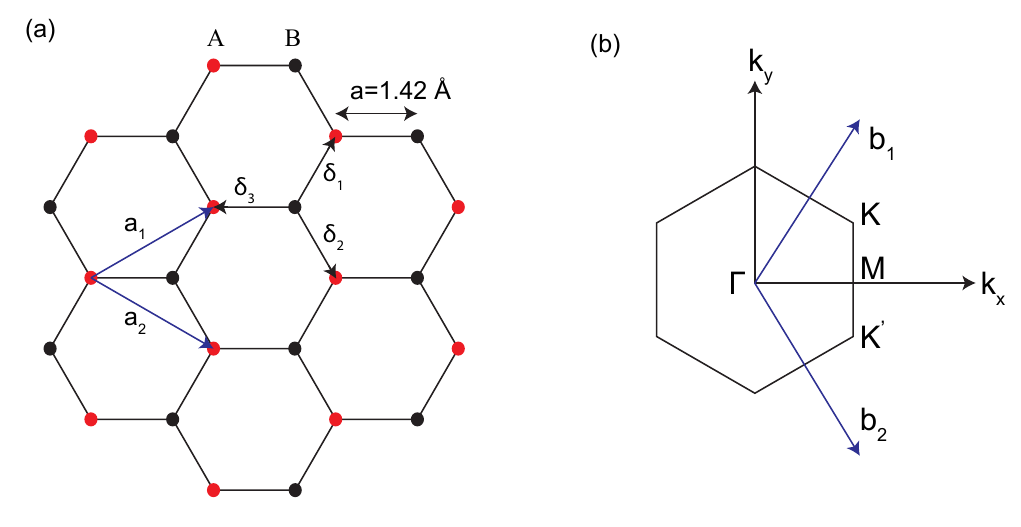}
 	\caption{\textbf{(a)} Carbon atoms are arranged in a honeycomb lattice structure. It can be thought of as two interpenetrating triangular lattices. Two nonequivalent sites are named $A$ (red) and $B$ (black), separated by distance $a = 1.42 $\AA. $\mathbf{a}_1$ and $\mathbf{a}_2$ are the lattice vectors and $\delta _i$, $i = 1,2,3$ are the nearest neighbor vectors. \textbf{(b)} $\mathbf{b}_1$ and $\mathbf{b}_2$ are the reciprocal lattice vectors. The region bounded by the hexagon boundary lines is the first Brillouin zone. The gap between the conduction and valence band vanishes at the Dirac point lattice vectors  $\mathbf{K}$ and $\mathbf{K}^{'}$. \textit{The figure is adapted from Saurabh Kumar Srivastav's PhD Thesis (2022). Reprinted with permission from the Indian Institute of Science\cite{srivastav2022quantized}}.}
 	\label{fig:SLG_lattice}
 \end{figure}

In single-layer graphene, carbon atoms are arranged in a honeycomb lattice structure as shown in ~\ref{fig:SLG_lattice} (a). To describe its lattice vectors, one can think of it as a triangular lattice with two sublattice basis A and B. The unit lattice vectors can be written as  

\begin{align}
\mathbf{a}_1 = \frac{a}{2}\left(3,\sqrt{3}\right) \, , \qquad
\mathbf{a}_2 = \frac{a}{2}\left(3,-\sqrt{3}\right) \, .
\label{eq:lattice_vector}
\end{align}

where $a = 1.42$ \AA ~is the carbon-carbon bond length. The corresponding reciprocal lattice vectors are given by 

\begin{align}
\mathbf{b}_1 = \frac{2\pi}{a}\left(1,\sqrt{3}\right) \, , \qquad
\mathbf{b}_2 = \frac{2\pi}{a}\left(1,-\sqrt{3}\right) \, .
\label{eq:re_lattice_vector}
\end{align}

The first Brillouin zone (BZ) of the graphene lattice is shown in Fig.~\ref{fig:SLG_lattice}(b). There are six corners in the first BZ, out of which only two points, known as `Dirac points', $\mathbf{K}$ and $\mathbf{K}^{'}$ are non-equivalent. The positions of $\mathbf{K}$ and $\mathbf{K}^{'}$ points in momentum space are given by
\begin{align}
\mathbf{K} = \frac{2\pi}{3a}\left(1,1/\sqrt{3}\right) \, , \qquad
\mathbf{K'} = \frac{2\pi}{3a}\left(1,-1/\sqrt{3}\right) \, .
\label{eq:Co_Dirac_point}
\end{align}

SLG's low energy band structure can be calculated using the tight-binding approach. For the scope of this review, we account for only the first three nearest neighbour sites, which are 
\begin{align}
\boldsymbol{\delta_1} = \frac{a}{2}\left(1,\sqrt{3}\right) \, , \qquad
\boldsymbol{\delta_2} = \frac{a}{2}\left(1,-\sqrt{3}\right) \, , \qquad
\boldsymbol{\delta_3} = -a\left(1,0\right) \, .
\label{eq:Co_nn}
\end{align}

The tight binding Hamiltonian accounting only the first three NN can be written in the basis of the wave-function amplitudes on the two sublattices $\Phi_{A}$ and $\Phi_{B}$, as follow;

\begin{align}
\widehat{H}_{SLG} = -\gamma_{0}\left(\begin{array}{cc} 0 & f(\mathbf{k}) \\ f^{*}(\mathbf{k}) & 0 \end{array}\right) 
\label{eqn1:SLG_Landau_Level_1}
\end{align}
where $\gamma_{0}=2.8$ eV is the nearest neighbour hopping between site $A\leftrightarrow B$. and 

\begin{align}
f(\mathbf{k}) = e^{-i\boldsymbol{k.\delta_{1}}}+e^{-i\boldsymbol{k.\delta_{2}}}+e^{-i\boldsymbol{k.\delta_{3}}}
\label{eqn1:f(k)_define}
\end{align}
The resulting eigenvalues of the Hamiltonian(\ref{eqn1:SLG_Landau_Level_1}) have two eigenvalues, which become identically zero at $\mathbf{K}$ and $\mathbf{K}^{'}$ points in the Brillouin zone. In most cases, only the low-energy bands around these points ($\mathbf{K}$ and $\mathbf{K}^{'}$), known as Dirac points and Valley in the Brillouin zone, are important. The low-energy Hamiltonian around these Dirac points can be written as 

\begin{align}
\widehat{H}_{SLG} = \hbar v_{F}\left(\begin{array}{cc} 0 & \xi k_{x}-ik_{y}\\ \xi k_{x}+ik_{y} & 0 \end{array}\right) 
\label{eqn1:gr_SLG_mat_4}
\end{align}
where $v_{F} = 3\gamma_{0}a/2\hbar \simeq 1\times 10^{6}$~m/s and $\xi=\pm1$ for $\mathbf{K/K^{'}}$ point.
The eigenvalues of the Hamiltonian(\ref{eqn1:gr_SLG_mat_4}) around $\mathbf{K}$ point is given by;
\begin{align}
E_{\pm}(k) = \pm \hbar v_{F} |\mathbf{k}|
\label{eqn1:linear_disp}
\end{align}

The resulting low-energy band dispersion of graphene has a four-fold degeneracy, 2 for the spin flavour and 2 for the Valley flavour. In the presence of a magnetic field perpendicular to the graphene plane $\mathbf{B} = B\hat{\mathbf{z}}$, the band dispersion gets modified; Following the minimal coupling approach of the vector potential $\mathbf{A} = (-By,0,0)$; the Hamiltonian(\ref{eqn1:gr_SLG_mat_4}) around $\mathbf{K}$ point can be written in terms of the defined operator $\pi \rightarrow (p_x-eBy+ip_y = -i\hbar \partial_x -eBy +\hbar \partial_y)$ and $\pi^{\dagger} \rightarrow (p_x-eBy-ip_y = -i\hbar \partial_x -eBy -\hbar \partial_y)$.
The resulting Hamiltonian in the magnetic field near the Dirac point $\mathbf{K}$ is given by

\begin{align}
H_{SLG,\mathbf{K}} = v_{F}\left(\begin{array}{cc} 0 &  \pi^{\dagger}\\ \pi & 0 \end{array}\right)
\label{eqn1:SLG_Landau_Level_13}
\end{align}
Operator $\pi$ and $\pi^{\dagger}$ act as a creation and annihilation operator respectively, such that 

\begin{align}
\pi \psi_N = -\frac{\sqrt{2}i\hbar}{l_B}\sqrt{N}\psi_{N-1},\\
\pi^{\dagger} \psi_N = \frac{\sqrt{2}i\hbar}{l_B}\sqrt{N+1}\psi_{N+1},\\ \pi \psi_0  = 0
\label{eqn1:SLG_Landau_Level_16}
\end{align}
Here, $\psi_N (x,y)$ is the scalar orbital Landau level wavefunction;

\begin{align}
\psi_N (x,y) = \frac{1}{\sqrt{2^{N}(N!)\sqrt{\pi}}}e^{\frac{ip_{x}x}{\hbar}}H_N\bigg(\frac{y}{l_B}-l_{B}k\bigg)exp\bigg[-\frac{1}{2}\bigg(\frac{y}{l_B}-l_{B}k\bigg)^2\bigg]
\label{eqn1:SLG_Landau_Level_15}
\end{align}
where $l_B=\sqrt{\hbar/eB}$ is the magnetic length scale, $N$ is the Landau level index and $H_N$ is $N^{th}$ Hermite polynomial. The resulting eigenvalues are given by,

\begin{align}
     \epsilon_{N} = sign(N) \frac{\sqrt{2}\hbar v_{F}}{l_B}\sqrt{|N|} = sign(N) \sqrt{2\hbar v_F^{2}eB|N|}
    \label{eqn1:SLG_Landau_Level_eigen}
\end{align}
The eigenstates at $\mathbf{K}$ point are given by
\begin{align}
    \forall N\geq1 : \Phi_{N,\pm} =\frac{1}{\sqrt{2}} \left(\begin{matrix}\Psi_N\\\mp i\Psi_{N-1}\\\end{matrix}\right) \\
    N=0: \Phi_{0} =\frac{1}{\sqrt{2}} \left(\begin{matrix}\Psi_0\\0\\\end{matrix}\right)
    \label{eqn1:SLG_Landau_Level_18}
\end{align}
whereas near $\mathbf{K'}$ point, eigenstates are
\begin{align}
    \forall N\geq1 : \Phi_{N,\pm} =\frac{1}{\sqrt{2}} \left(\begin{matrix}\pm i\Psi_{N-1}\\\Psi_N\\\end{matrix}\right) \\
    N=0: \Phi_{0} =\frac{1}{\sqrt{2}} \left(\begin{matrix}0\\\Psi_0\\\end{matrix}\right)
    \label{eqn1:SLG_Landau_Level_19}
\end{align}
\begin{figure}[t!]
 	\centering
 	\includegraphics[width=1.0\textwidth]{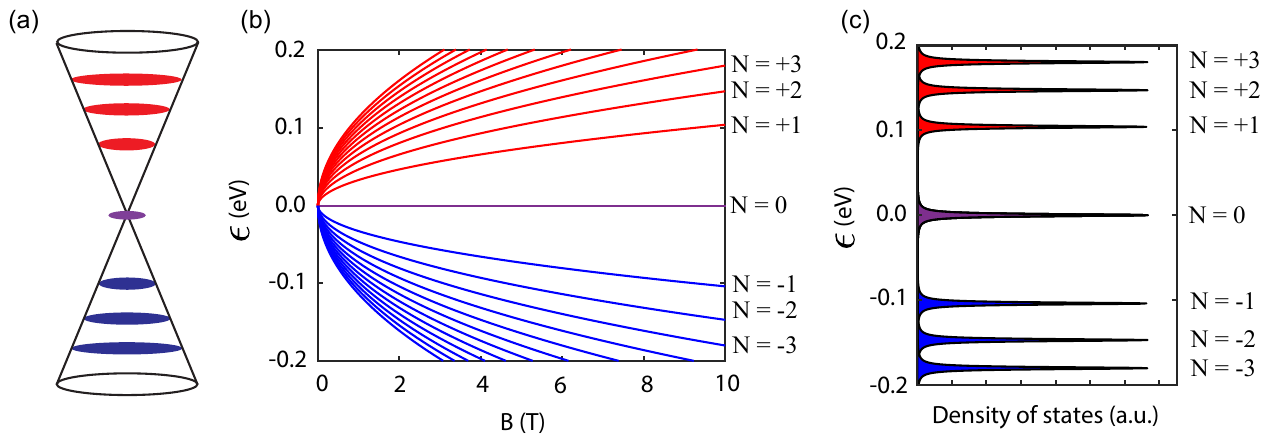}
 	\caption{\textbf{(a)} In the presence of a magnetic field, Landau level forms at discrete energy intervals. \textbf{(b)} Landau level spectrum is plotted for both conduction and valence band as a function of the magnetic field $\mathbf{B}$ for different Landau Level index $N$. Here, Landau level energy disperses as $\epsilon = sgn(N)\sqrt{2\hbar v_F^{2}eB|N|}$ and as a result energy gap reduces as one approaches higher Landau levels. It is in sharp contrast to conventional two-dimensional electron gas, where the energy gap remains the same irrespective of the Landau level index. The presence of zero energy Landau level, equally shared by electron and hole, also makes SLG a unique two-dimensional material. \textbf{(c)} The corresponding density of states of Landau levels after incorporating the Lorentzian broadening of the $3 meV$ for each Landau level for $B=10$ T. \textit{The figure is adapted from Saurabh Kumar Srivastav's PhD Thesis (2022). Reprinted with permission from the Indian Institute of Science\cite{srivastav2022quantized}}.} 
 	\label{fig:SLG_Landau_level}
 \end{figure}

It should be noted that for all Landau level $N\geq1$, the electronic density is shared by both sublattice site $A$ and $B$. However, for $N=0$ state, near $\mathbf{K/K'}$ valley, electronic density only resides on the sublattice $A/B$. The resulting Landau level energy spectrum and the density of states are plotted in Fig.~\ref{fig:SLG_Landau_level}(b) and (c), respectively. The presence of the zero energy Landau level (ZLL) and four-fold spin valley degeneracy of SLG, the Hall conductivity $\sigma_{xy}$ exhibits an unconventional sequence of the quantization described by\cite{girvin1987quantum,macdonald1989quantum,gusynin2005unconventional,herbut2007theory,ostrovsky2008theory,yang2007spontaneous,jiang2007quantum,novoselov2007room} 
 \begin{equation}
    \sigma_{xy} = \pm 4 \bigg(N+\frac{1}{2}\bigg)\frac{e^2}{h}
    \label{eqn1:SLG_Landau_Level_20}
\end{equation}
It should be noted that the Landau level's wavefunction in SLG differs significantly from the conventional \ch{GaAs/AlGaAs}. First, here, for SLG, the wavefunction of the $N^{th}$ Landau level has the orbital component of the $N^{th}$ and $N^{th}-1$ conventional Landau level, except ZLL. This difference can lead to different values of Haldane pseudopotentials for the LLs with different orbital structures and hence affects the sequence of the FQH states in different LLS. The second major difference, which is more pronounced in ZLL is the different sublattice structure of wavefunctions near degenerate $\mathbf{K}$ and $\mathbf{K'}$. As a consequence, it affects the valley and or spin symmetry-breaking of ZLL and the ground state of $\nu=0$ insulating state.

\subsection{Symmetry breaking and quantum Hall Ferromagnetism}
\begin{figure}[t!]
 	\centering
 	\includegraphics[width=1.0\textwidth]{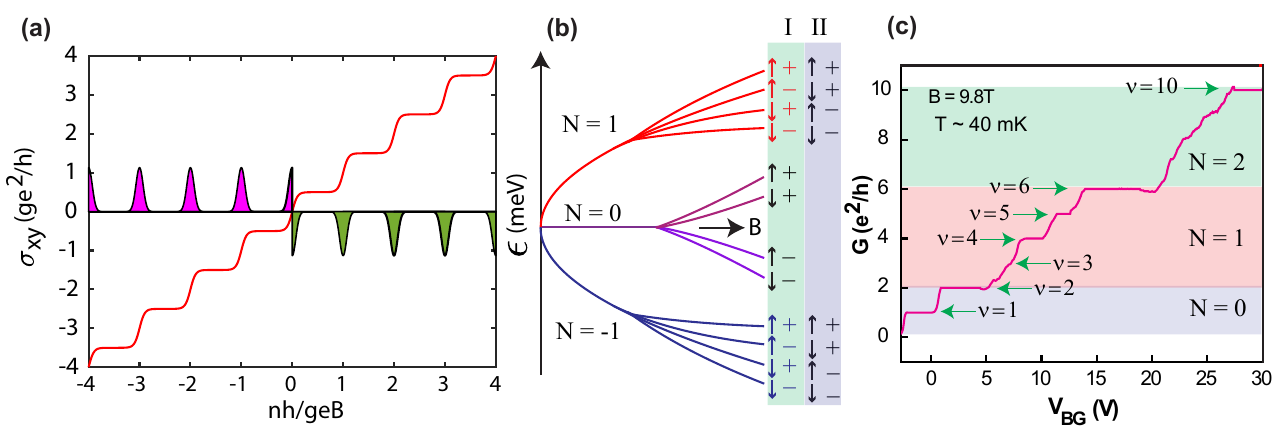}
 	\caption{\textbf{(a)} In absence of any interaction, the single particle spectrum of Landau levels gives the quantum Hall plateaus at $g\big(N+\frac{1}{2}\big) \frac{e^2}{h}$, where $g = 4$ is the four fold spin-valley degeneracy and $N$ is the Landau level index. The distance
     between the quantum Hall plateaus along the density axis is defined by the density of states $\frac{gB}{\phi_0}$ on each Landau level, which remarkably does not depend on the zero field spectrum of the system. Here, $B$ is the magnetic field and $\phi_0 = \frac{h}{e}$ is the flux quantum. In addition to the plateaus in transverse conductance, the corresponding sequence of the Landau levels are shown in magenta and green for holes and electrons, respectively. \textbf{(b)} However, in presence of interactions and high magnetic field, the four-fold spin valley degeneracy breaks, which leads to the observation of the plateaus at the integer values outside the conventional sequence. The schematic shows the energy spectrum for the spin (I) or valley(II) polarization for the Landau levels after spin-valley symmetry breaking. Two different possible scenarios associated with two sequences of edge states are labelled by $I$ and $II$, respectively,
     In the schematic, the edge states’ spin and valley
     polarization are denoted by $\uparrow /\downarrow$ and $\pm$, respectively .\textbf{(c)} Transverse conductance is plotted as function of the back gate voltage at $B = 9.8$ T and $40$ mK of bath temperature for a single layer graphene device. In addition to the conventional plateaus at $2\frac{e^2}{h}, 6\frac{e^2}{h}, 10\frac{e^2}{h}$, one can clearly observe the well developed symmetry broken plateaus of $N = 0$ and $N = 1$ Landau levels. \textit{The figure is adapted from Saurabh Kumar Srivastav's PhD Thesis (2022). Reprinted with permission from the Indian Institute of Science\cite{srivastav2022quantized}}.} 
 	\label{fig:SLG_QH_sequence}
 \end{figure}

Along with the unconventional sequence of the QH plateaus due to the $\pi$ Berry's phase, the spin and valley degeneracy makes graphene a wonderful system for observing the rich physics associated with the multicomponent quantum Hall phases\cite{nomura2006quantum,young2012spin,yang2007spontaneous,zhang2006landau,young2014tunable,abanin2006spin,amet2014selective,sheng2006quantum,alicea2007interplay,goerbig2011electronic,goerbig2006electron,peres2005coulomb,yang2006collective,PhysRevB.98.155421}. In SLG, the relevant energy scales are the Landau quantization energy $\epsilon_{N}$ (cyclotron energy) and the long-range Coulomb interaction $E_{C}=\frac{e^2}{\epsilon l_{B}}$. At low magnetic fields, $\epsilon_{N}$ and $E_{C}$ do not depend on the spin or valley degrees of freedom. So the combined spin-valley flavour degeneracy can be described in terms of a single approximate $SU(4)$ isospin symmetry which corresponds to the invariance of long-range electron-electron $(E_{C})$ interaction under a rotation within the fourfold spin/valley internal space. However, the long-range Coulomb interaction $(E_{C})$ can drive the system through a ferromagnetic instability, in which the order parameter corresponds to a finite polarization in a specific direction within the $SU(4)$ isospin space \cite{nomura2006quantum,young2012spin,yang2007spontaneous,zhang2006landau,young2014tunable,abanin2006spin,amet2014selective,sheng2006quantum,alicea2007interplay,goerbig2011electronic,goerbig2006electron,peres2005coulomb,yang2006collective}. At the integer fillings within a partially filled quartet Landau levels, this order parameter is predicted to lead to a finite
gap for charged excitations and a robust quantum Hall effect for integers outside the sequence described in Eqn.~\ref{eqn1:SLG_Landau_Level_20}. The exact $SU(4)$ polarization for a given sample depends on the interplay between the anisotropies arising from the $SU(4)$ symmetry-breaking term like Zeeman splitting, lattice scale interactions and even the disorder (that might not break the $SU(4)$ symmetries). Fig.~\ref{fig:SLG_QH_sequence}(c) shows the plot of transverse conductance as a function of the back gate voltage at $B = 9.8$ T and $40$ mK of bath temperature for a single layer graphene device encapsulated between the two hexagonal boron nitride substrate (hBN). In addition to the conventional plateaus at $2\frac{e^2}{h}, 6\frac{e^2}{h}, 10\frac{e^2}{h}$, well-developed symmetry broken plateaus of $N = 0$ and $N = 1$ LLs are also seen.

\section{Fractional quantum Hall effect in single-layer graphene}
At the partial filling of LLs, when the kinetic energy is completely quenched, electron-electron interaction leads to the emergence of fractional quantum Hall (FQH) states. Due to the approximate SU(4) symmetry of SLG, the symmetry breaking at partial filling of LLS is quite complex. It can lead to a condition where multiple possible orders, such as single-component and multi-component FQH states, compete with each other for the ground state. The experimental appearance or absence of the FQH sequences in SLG can be mainly explained either via a single-component or multi-component composite fermion (CF) model. Conventionally, in single component CF model, each electron captures an even number, say 2p, of the quantized vortices(or flux quanta) and turns itself into weakly interacting topological particles, known as composite fermions (CF). Due to the vortex attachment, weakly interacting CFs experience an effective magnetic field $B^{*}$, which is related to the externally applied field $B$ via a relation\cite{jain1989composite,jain2007composite};
\begin{align}
    B^{*} = B-2p\rho \phi_0
    \label{eqn1:CF_0}
\end{align}
where $\rho$ is the electron density, $\phi_0 = h/e$ is the flux quantum and 2p is an even integer. Equivalently, one can write that electrons at filling factor $\nu$ convert into a composite fermion with filling $\nu^{*}$, related via 
\begin{align}
\nu = \frac{\nu^{*}}{2p\nu^{*} \pm 1}
    \label{eqn1:CF_1}
\end{align}
This single-component non-interacting CF model accurately describes the sequence and hierarchy of FQH states observed between $0<\nu<1$ of SLG\cite{PhysRevLett.122.137701,zibrov2018even,scarola2001phase}. However, FQH states observed experimentally between $1<\nu<2$ raise concern over the applicability of single component CF model in this branch of symmetry-broken LLs\cite{dean2011multicomponent,feldman2012unconventional,feldman2013fractional}. Particularly, the absence of $\nu=5/3$ was quite surprising. Since $5/3=2-1/3$ is closest to the conventional $1/3$ state, naively one would expect to observe well-developed FQH states at $\nu=5/3$. Furthermore, the emergence of only even numerator states with a step of 2 between the consecutive FQH states in $1<\nu<2$, suggests the partial lifting of four-fold spin and valley and a transition into an approximate SU(2) symmetry phases. Careful observation of the emergent FQH states suggests that these responses have close analogies with the IQH effect of real electrons observed in SLG with a four-fold degeneracy of LLs. Within the CF picture, these states are usually described via a two-component CF model, similar to what has been used to explain the sequence of multicomponent FQH states observed in \ch{AlAs} and strained silicon\cite{padmanabhan2010ferromagnetic,lai2004fractional}. To date, many properties of the ground states and the excitation of FQH states in graphene are unknown and demand a detailed and systematic investigation.

\section{Topological order and its Detection in integer and fractional quantum Hall states}
In the topological phase of matter, some of the physical properties remain insensitive to the local perturbations of the system\cite{wen2005introduction,levin2006detecting,kitaev2006topological}. These physical properties may include fractionalization\cite{wen2005introduction}, long-range quantum entanglement\cite{kitaev2006topological}, and topological degeneracies in the energy spectrum of the system\cite{wen2005introduction,levin2006detecting,kitaev2006topological}. Mathematically, the topological order of a quantum phase is usually characterized by some of the topological invariant numbers. Topological invariant numbers are often used to characterize the class of topological order of phase that has protected gapless edge modes when the system is kept on a space with the boundary\cite{wen2005introduction,levin2006detecting,kitaev2006topological}. The integer and fractional quantum Hall phases are the first set of the discovered topologically ordered phases that have protected gapless edge modes at their boundary. Usually, the knowledge of the topological order of the IQH and FQH states demands a detailed understanding of the charge and statistics of the anyons, which exist in the gapped bulk of the 2D electron liquid. However, accessing the bulk of the 2D electron liquid in the experiment is challenging. Thanks to the bulk-edge correspondence principle, the topological order of the gapped bulk of IQH and FQH phases can be determined by examining the physics of the gapless edge modes. 

\subsection{Edge states in Integer and fractional quantum Hall states}
\label{subsection:FQH_edge}
\begin{figure}[t!]
 	\centering
 	\includegraphics[width=1.0\textwidth]{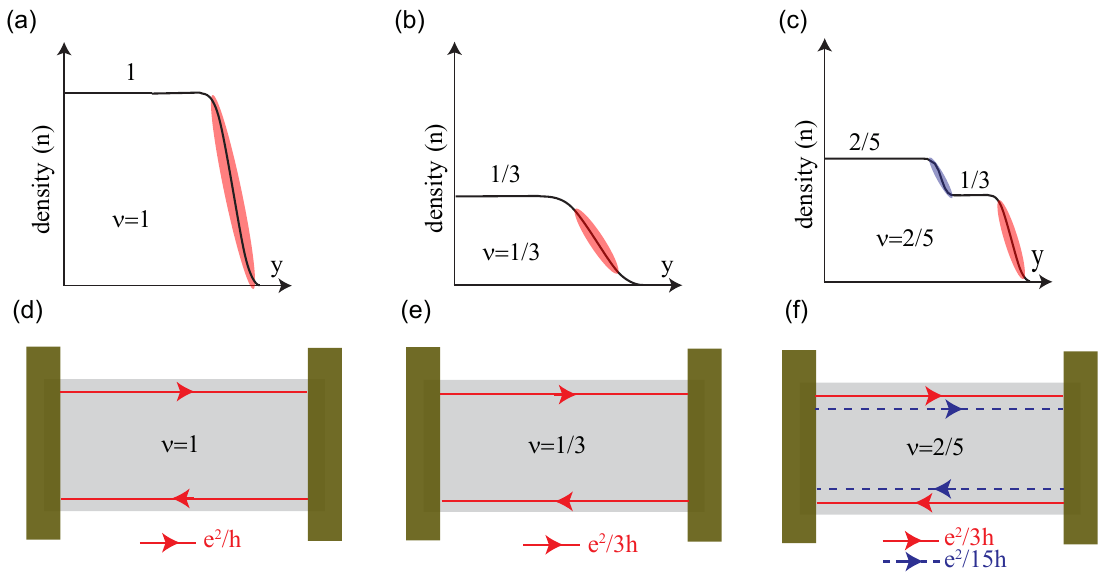}
 	\caption{\textbf{(a)} charge density profile of $\nu = 1$, \textbf{(b)} $\nu = 1/3$,  and \textbf{(c)} $\nu = 2/5$ states near the edge. In these states, charge density and potential profile drops monotonically to zero. The schematic of the downstream edge modes at the physical boundary of the sample is shown in \textbf{(d)},\textbf{(e)}, and \textbf{(f)} for $\nu = 1, 1/3,$ and $\nu = 2/5$ states, respectively. The conductance of downstream edge modes in integer ($\nu = 1$) and $\nu = 1/3$ particle-like states are $e^2/h$ and $e^2/3h$, respectively. For $\nu = 2/5$ state, two downstream edge modes with conductance $e^2/15h$ (shown with blue dashed colour: inner mode) and $e^2/3h$ (shown with red colour: outer mode), respectively.\textit{The figure is adapted from Saurabh Kumar Srivastav's PhD Thesis (2022). Reprinted with permission from the Indian Institute of Science\cite{srivastav2022quantized}}.} 
 	\label{fig:particle_like_edge_state}
 \end{figure}

\begin{figure}[t!]
 	\centering
 	\includegraphics[width=1.0\textwidth]{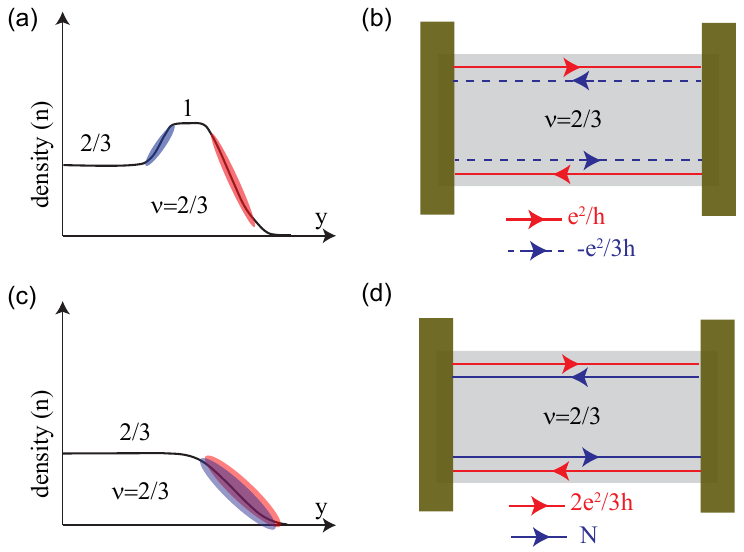}
 	\caption{\textbf{(a)} charge density profile of $\nu = 2/3$ at the sharp edge. \textbf{(b)} A schematic of the bare edge modes, with conductance $e^2/h$ (downstream mode shown with red colour: outer mode) and $e^2/3h$ (upstream mode shown by blue dashed colour: inner mode), respectively. \textbf{(c)} charge density profile of $\nu = 2/3$ at a disordered edge. In the presence of the disordered dominated tunnelling between the counter-propagating bare edge modes, one gets a decoupled charge mode of conductance $2e^2/3h$ and an upstream neutral mode, which does not carry any charge but can support the energy transport. \textbf{(d)} The schematic of the decoupled downstream charged mode with conductance $2e^2/3h$ (shown with red colour: outer mode) and upstream neutral mode (shown with blue colour: inner mode). \textit{The figure is adapted from Saurabh Kumar Srivastav's PhD Thesis (2022). Reprinted with permission from the Indian Institute of Science\cite{srivastav2022quantized}}.} 
 	\label{fig:neutral_mode_in_2_by_3}
 \end{figure}

Similar to the IQH case (Fig.~\ref{fig:particle_like_edge_state}(a)), the charge density for the particle-like FQH states $(\nu<1/2)$ drops monotonically from its bulk values to zero as one approaches the edge from the bulk\cite{macdonald1990edge}, as shown in Fig.~\ref{fig:particle_like_edge_state}(b, c) for $\nu = 1/3, 2/5$. It leads to the emergence of the one-dimensional downstream edge modes near the physical boundary of the sample\cite{macdonald1990edge}, as shown in Fig.~\ref{fig:particle_like_edge_state}(e, f) for $\nu = 1/3, 2/5$. The applied magnetic field's direction dictates the edge modes' downstream chirality\cite{halperin1982quantized,beenakker1990edge}. However, the charge density profile does not decrease monotonically for the hole-conjugate FQH states $(1/2<\nu<1)$. For example, at $\nu =2/3$, the filling factor increases from $2/3$ to $1$ and then drops back to zero\cite{macdonald1990edge}. In other words, $2/3$ liquids can be considered an FQH state of $\nu = 1/3$ holes in the IQH state at $\nu = 1$ electron liquid. As a consequence, two counter-propagating edge modes of charge $e$ with conductance $e^2/h$ and charge $-e/3$ with conductance $e^2/3h$ emerge at the sample boundary\cite{macdonald1990edge,johnson1991composite}. Fig.~\ref{fig:neutral_mode_in_2_by_3}(a, b) shows the schematics of charge density profiles and edge states of the $\nu = 2/3$ FQH state. However, in the presence of the disorder-dominated random inter-edge tunnelling between, Kane-Fisher-Polchinski\cite{Kane1994} found a decoupled charge mode of conductance $2e^2/3h$ and a neutral mode, which does not carry any charge but can carry the energy. The schematics of such normalized edge modes are shown in Fig.~\ref{fig:neutral_mode_in_2_by_3}(c, d). 

Since the neutral modes do not couple with the charged excitations of the external probes, it becomes challenging to detect them via the conventional electrical conductance measurement. Although several proposals were made to see the neutral modes which involve the measurement of the tunnelling exponent in constrictions\cite{kane1995impurity}, looking for resonance in long constriction\cite{overbosch2009long}, and looking for the heating effects on shot noise\cite{feldman2008charge,grosfeld2009probing}, the first experimental proof was demonstrated by Bid et al.;\cite{bid2010observation} for filling factor $\nu =2/3, 3/5,$ and $5/2$ in \ch{GaAs/AlGaAs} sample. This experiment involved an upstream quantum point contact (QPC) constriction from an energized ohmic contact. An excited neutral mode emanates from the `hot spot' at the source, propagates upstream along the edge, and impinges on a partially pinched QPC. This leads to observed current fluctuations. Recently, upstream neutral modes for $\nu =2/3, 3/5$ mode have been also observed in bilayer graphene devices\cite{kumar2022observation}.

\subsection{Detection of topological order}
Quantized electrical and thermal Hall conductance in quantum Hall states have been known for a long time back theoretically\cite{kane1996thermal,kane1997quantized} since their first set of experimental observations. Although electrical Hall conductance has been widely used to understand the topological order of a quantum Hall state, it is insufficient in the hierarchical fractional quantum Hall states, where the edge structure is complicated, and transport may occur via the downstream and upstream modes. The electrical Hall conductance only reveals the number and the conductance of the downstream charged chiral edge modes. Still, it is independent of the edge modes' total number, chirality, and character. By contrast, the quantized thermal Hall conductance is not only sensitive to the downstream charged modes, it can also detect the other upstream modes, including the chargeless neutral modes, which are not detectable in electrical Hall conductance measurement\cite{kane1996thermal,kane1997quantized,cappelli2002thermal}. 

Similar to the quantization of electrical conductance, the quantized thermal conductance of a single ballistic channel only depends on the fundamental constants of nature and is given by $\kappa_{0} T = \frac{\pi^2 k_{B}^2}{3h}T$, where $k_B$ and $h$ are the Boltzmann's and Planck's constant, respectively. If multiple downstream edge modes exist, such as IQH and particle-like FQH states, quantized thermal conductance becomes $G_{Q} = N_d\kappa_{0}T$, where $N_d$ is the total number of downstream edge modes. However, the situation becomes complex in the inter-mode tunnelling between counter-propagating edge modes at certain fillings, such as for $2/3$ and $3/5$, where some chiral edge modes propagate in the upstream direction. In 1997, Kane and Fisher derived the quantized thermal Hall conductance for the abelian fractional quantum Hall states for an ideal impurity-free edge\cite{kane1997quantized}. For the FQH states with counter-propagating edge modes, their analysis assumes a substantially longer edge propagation length than the thermal equilibration length. They found that quantized thermal Hall conductance at any filling of abelian FQH state is given by\cite{kane1997quantized} $G_{Q} = (N_d - N_u)\kappa_{0}T$,
where $N_d$ and $N_u$ are the numbers of the downstream and upstream edge modes, respectively. However, if the edge propagation length is smaller than the thermal equilibration length $(l_H)$, thermal Hall conductance takes the value of $G_{Q} = (N_d + N_u)\kappa_{0}T$. Since quantized $G_{Q}$ depends on the $N_d$ and $N_u$, it can also be used as a powerful experimental tool to detect the upstream neutral modes and hence the exact topological order of complex FQH states, which were not possible in electrical conductance measurements. Furthermore, in contrast to quantized electrical conductance, quantized thermal conductance remains independent of the statistics of the carriers. It remains the same for the fermions, bosons, anyons \cite{kane1997quantized,rego1999fractional} except in the case of the Majorana mode, where the quantized thermal conductance becomes half of its quantum limit,\cite{read2000paired,sumiyoshi2013quantum,nomura2012cross,kasahara2018majorana,yokoi2021half}. It makes the thermal conductance measurement a powerful technique which can be used to distinguish non-abelian order of even-denominator FQH states. Usually, the non-abelian order of FQH state, such as $5/2$ is believed to host the Majorana modes in its edge structure\cite{moore1991nonabelions,morf1998transition,storni2010fractional,rezayi2017landau,levin2007particle,lee2007particle,wen1991non,yang2013influence,yang2014experimental,son2015composite,zucker2016stabilization,fidkowski2013non}. Hence the observation of the half-quantized thermal conductance will be the smoking-gun evidence of the Majorana mode. Experimentally, the half quantum of the thermal conductance has been measured at $5/2$ FQH state in GaAs/AlGaAs based two-dimensional electron gas\cite{banerjee2018observation}.

\subsection{Measurement of quantized thermal conductance in IQH and FQH states}

\begin{figure}[t!]
 	\centering
 	\includegraphics[width=1.0\textwidth]{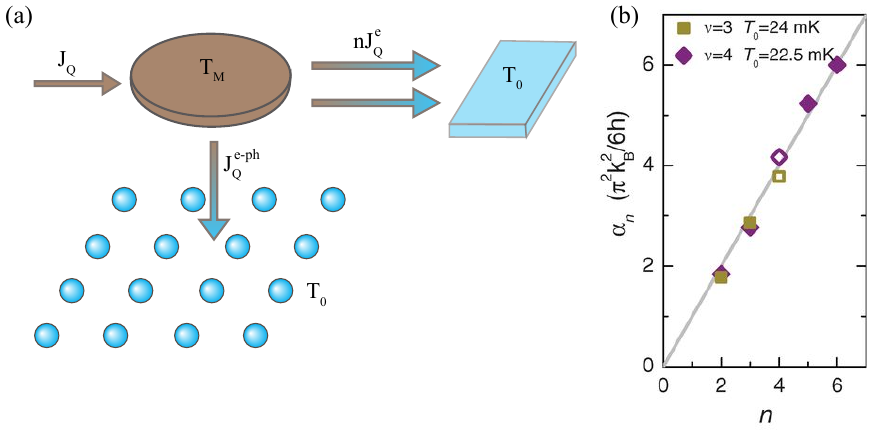}
 	\caption{\textbf{(a)} Basic principle of thermal conductance measurement of QH states. A micrometre size metallic floating contact is heated to $T_M$ by injecting a known Joule power $J_Q$. The temperature of the floating contact is set by the heat balance equation $J_Q = nJ_Q^e + J_Q^{e-ph}$, where $nJ_Q^e$ is the electronic contribution to heat flow via $n$ ballistic QH edge channels from floating contact to cold bath at temperature $T_0$ and $J_Q^{e-ph}$ is the transfer of heat from the hot electrons in floating contact toward the cold phonon bath at temperature $T_0$. \textbf{(b)} Extracted electronic heat current factor $\alpha_n$ (shown by symbols) defined as $nJ_Q^e/(T_M^2-T_0^2)$ normalized by $\pi^2k_B^2/6h$ is plotted as a function of the total number of channel $n$. The grey line shows the predictions for the quantum limit of the heat flow. \textit{The figure is adapted from Jezouin et al. (2013). Reprinted with permission from AAAS}.} 
 	\label{fig:heat_flow_scheme}
 \end{figure}

Different from electrical conductance measurement, quantized thermal conductance measurement is complicated and tricky. It is evident from the fact that although the quantization of the electrical and thermal conductance of the IQH and FQH states has been known for decades, the first experimental measurement of quantized thermal conductance of quantum Hall states was reported 33 years after the first experimental demonstration of the IQH states in the pioneering work of Jezouin et al. \cite{jezouin2013quantum} in GaAs/AlGaAs based two-dimensional electron gas. The basic principle of the experimental set-up is shown in Fig.~\ref{fig:heat_flow_scheme}(a). In this measurement scheme, a micrometer-sized metallic floating reservoir is connected to the cold electronic reservoir via an adjustable number $n$ of the ballistic QH edge channels and a phonon cold bath at temperature $T_0$. The electron temperature of the floating metallic reservoir is heated to a temperature $T_M$ by injecting a known joule heating power $J_Q$ to it. The temperature of the floating reservoir is determined by a heat balance equation given by;

\begin{equation}
    J_Q=J_{Q}^{e} + J_{Q}^{e-ph} = 0.5N\kappa_{0}(T_M^2-T_0^2) + J_{Q}^{e-ph}
    \label{eq:1}
\end{equation}

Here, the first and the second terms in this heat balance equation correspond to the electronic contribution to heat flow via $N$ ballistic edge channel and heat loss due to electron-phonon coupling from the floating reservoir to the phonon cold bath. The central result of Jezouin et al. is shown in Fig.~\ref{fig:heat_flow_scheme}(b), where $\alpha_n$ measured electronic thermal conductance normalized by $0.5\kappa_{0}$ is plotted as a function of the $n$ number of the ballistic edge channel. The unit slope of the experimental fit data corresponds to the theoretically expected quantized limit on the thermal conductance.  Later, a similar measurement was extended to the FQH regime by Banerjee et al. \cite{banerjee2017observed,banerjee2018observation} in the same system. Banerjee et al. \cite{banerjee2017observed} demonstrated the universality of the quantized thermal conductance for the anyonic heat flow. They further reported the observation of a half-integer of the quantized thermal conductance for even-denominator FQH state $5/2$ \cite{banerjee2018observation}. The measured thermal conductance value of $2.5 \kappa_0 T$ was interpreted as the particle–hole Pfaffian (PH-Pfaffian) topological order of $5/2$ state. As mentioned above, the experiments by Jezouin et al. and Banerjee et al. were performed in \ch{GaAs/AlGaAs} based two-dimensional electron gas.

\begin{figure}[t!]
 	\centering
 	\includegraphics[width=0.75\textwidth]{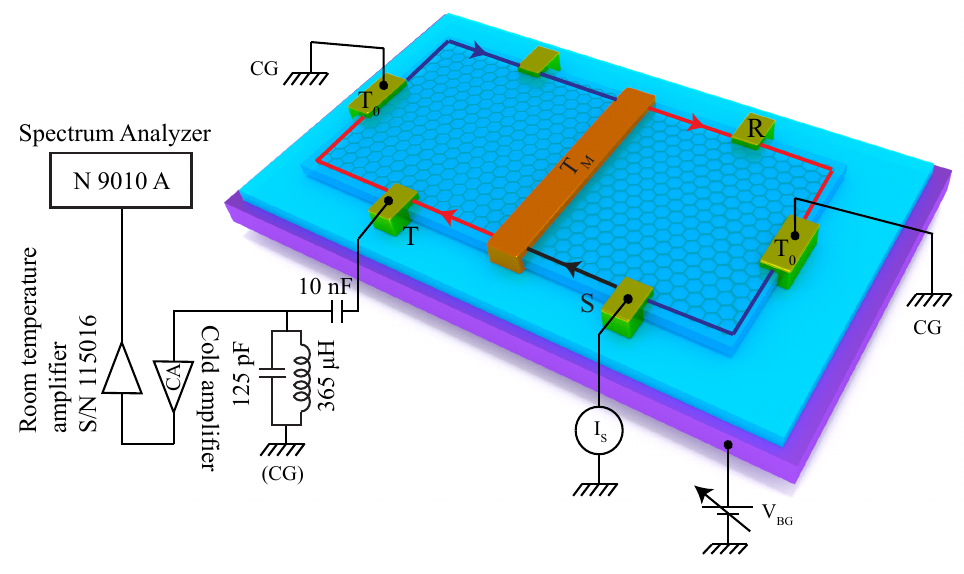}
 	\caption{Schematic of the graphene device with thermal conductance measurement setup. The device is set in the integer QH regime at filling factor $\nu = 1$, where one chiral edge channel (line with arrow) propagates along the edge of the sample. The current $I_S$ is injected (black line) through the contact `S', which is absorbed in the floating reservoir (orange contact). The chiral edge channel (red line) at potential $V_M$ and temperature $T_M$ leave the floating reservoir and terminates into two cold grounds (CGs). The blue lines show the cold edges (without current) at temperature $T_0$. The resulting increase in the electron temperature $T_M$ of the floating reservoir is determined from the measured excess thermal noise at contact $T$. A resonant (LC) circuit, situated at contact $T$, with resonance frequency $f_0 \approx 750$ kHz, filters the signal, which is amplified by the cascade of amplification chain (cryogenic (cold) preamplifier placed at 4K plate and a room temperature amplifier). A ceramic capacitance of 10 nF was introduced between the sample and inductor to block the DC current along the measurement line. Last, the amplified signal is measured by a spectrum analyzer. \textit{The figure is adapted from Saurabh Kumar Srivastav's PhD Thesis (2022). Reprinted with permission from the Indian Institute of Science\cite{srivastav2022quantized}}.}
 	\label{fig:meas_setup}
 \end{figure}

Although The quantization of the electrical Hall conductance in graphene, a van der Waals material was reported in 2005 by A. Geim and Philip Kim group separately\cite{novoselov2005two,zhang2005experimental}, the first experiment of quantized thermal conductance in QH states in graphene was reported in 2019\cite{doi:10.1126/sciadv.aaw5798}. In this experiment, graphene was encapsulated between two hexagonal boron nitride substrates and was gated by either the \ch{SiO_2/Si} gate or the graphite back gate. In addition to having the Hall probe metallic contacts, a micron-size metallic floating contact was connected in the middle of the graphene flake via one-dimensional edge contact. It is important to emphasize that below the metallic floating reservoir, the graphene region was entirely etched, and after applying the sufficient magnetic field and the gate voltages, the QH edge channels directly terminate and leave the floating reservoir from one side of the graphene chamber to the other side. To heat the floating reservoir, a DC current $I_S$ was injected at contact `S' as shown in Fig. ~\ref{fig:meas_setup}. This setup helps to create a hot reservoir at milli-Kelvin temperature. The cold-grounded (CG) metal contacts, as shown in Fig. ~\ref{fig:meas_setup}, serve as a cold reservoir. Now, one has two reservoirs of different temperatures. To extract the thermal conductance, one needs to know the thermal current flowing between the two reservoirs via the electronic channel and the electron temperatures of the reservoir.

\subsection{Determination of the thermal current $J_Q$:}
 \label{j_Q_determination}
 The thermal current flowing between the two reservoirs can be easily estimated for integer and particle-like fractional quantum Hall states, where only downstream edge mode exists. In this scenario, all dissipated power near the floating contact is used entirely to heat the metallic floating contact. Hence, the thermal current flowing between two reservoirs will be exactly the same as the total dissipated power near the floating contact. A DC current $I_S$ injected at the source contact $S$ (Fig.~\ref{fig:meas_setup}) flows from the source contact $S$ to the floating reservoir along the chiral QH edge channels. 
 
 The outgoing current from the floating reservoir splits into two equal parts, each propagating along the outgoing chiral edge from the floating reservoir to the cold grounds. The floating reservoir reaches a new equilibrium potential $V_{M} = \frac{I_S}{2 \nu G_0}=\frac{V_S}{2}$ ( where $G_0 = e^2/h$) with the filling factor $\nu$ of graphene determined by the $V_{BG}$, whereas the potential of the source contact is $V_{S} = \frac{I_S}{\nu G_0}$. Thus, the power input to the floating reservoir is $P_{in} = \frac{1}{2}(I_S V_S) = \frac{I_S^2}{2 \nu G_0}$, where the pre-factor of $1/2$ results because half power dissipates at the back of source contact (\ref{fig:meas_setup}). Similarly, the outgoing power from the floating reservoir is $P_{out} = \frac{1}{2}(2 \times \frac{I_S}{2}V_M) = \frac{I_S^2}{4 \nu G_0}$. Thus, the resultant injected power dissipation in the floating reservoir due to joule heating is $J_{Q} = P_{in}-P_{out} = \frac{I_S^2}{4 \nu G_0}$.

 However, the situation may be complicated in hole-conjugate fractional quantum Hall states like $\nu = 2/3$, which also supports the upstream and downstream modes. The downstream and upstream modes can have different chemical potentials near the metallic floating contact and the source contact. Due to the charge tunnelling between the counter-propagating modes, the edge channel attains the equilibrium chemical potential for the propagation length larger than the charge equilibration length ($l_{eq}^{c}$). In this scenario, the total dissipated joule power also remains the same. Still, some portion is dissipated in the tunnelling regions outside the ohmic reservoirs, and the rest is dissipated in the floating metallic contact. In this case, one can not equate the total dissipated power to the thermal current $J_Q$. However, suppose one assumes that all heat from the charge equilibration region returns to the nearby ohmic reservoirs. In that case, one can safely use the formula derived for the integer and particle-like fractional quantum Hall states. Such an assumption is justified if the charge equilibration length is much smaller than the reservoir size. Fortunately, the charge equilibration length in a graphene-based system is expected to be quite small; hence, we can safely equate the total dissipated power near the floating contact to the thermal current $J_Q$. Here we would also like to define the characteristic length scales like charge and thermal equilibration length, which will be extensively used in this review.

\textbf{Charge equilibration length ($l_{eq}^{c}$):} Charge equilibration length ($l_{eq}^{c}$) is defined as the minimum propagation length required to achieve the complete voltage equilibration between the counter-propagating edge modes. In other words, all counter-propagating edge modes should attain the same electrostatic voltage after a propagation length of $l_{eq}^{c}$. 

\textbf{Thermal equilibration length ($l_{eq}^{H}$):} Thermal equilibration length ($l_{eq}^{H}$) is defined as the minimum propagation length required to achieve the same temperature between the counter-propagating edge modes. In other words, all counter-propagating edge modes should attain the same temperature profile after a propagation length of $l_{eq}^{H}$.
 
\subsection{Determination of the electron temperatures $T_M$:}
The resulting increase in the electron temperature $T_M$ of the floating contact was determined by measuring the excess thermal noise at contact `T'. To avoid the contribution from the 1/f noise, a resonant L//C tank circuit was used to shift the operating frequencies close to the 0.7 MHz regimes. The circuit consists of an inductor made of superconducting NbTi wire, and the capacitance developed along the coaxial line connecting the sample to the cryogenic amplifier. A ceramic capacitance of 10 nF was also introduced between the sample and inductor to block the DC current along the measurement line and isolate the DC voltage at the cryogenic amplifier gate port from the sample bias voltage. The output voltage from the cryogenic amplifier was further amplified using a room-temperature voltage amplifier. After the second amplification stage, the resulting signal was measured using a spectrum analyzer. The measured excess noise at contact `T' is related to the electron temperature $(T_M)$ of the floating contact via the relation,
\begin{equation}
    S_I=\nu k_B(T_M-T_0)
    \label{eq:2}
\end{equation}
where $\nu$ is the bulk filling of the QH state. Furthermore, it is worth emphasising the various conditions that need to be fulfilled to justify the use of noise thermometry for electron temperature $(T_M)$ determination.

\indent
\textbf{Continuous energy levels:} The energy level spacing of the micron size metallic floating contact must be negligible compared to the other relevant energy scales. Since, in typical graphene devices, edge contacts are mainly made via gold, one can easily estimate the energy level spacing. Using the density of states ($D_{E} \simeq 1.14\times10^{47}J^{-1}m^{-3}$), for the gold and the typical volume of the $\Omega \approx 0.5 \mu$m$^3$, the average energy level spacing becomes $\sim 1/(D_{E}\times \Omega)\sim k_B\times 1.45 \mu K$, which is much smaller than the typical electron temperature $8-40$ mK ranged achieved in thermal conductance measurement set-up. Hence, for micron-sized floating contact, one can safely ignore the issues related to the discrete energy levels of the floating metallic contact in our devices.

\textbf{Quasi-equilibrated electronic distribution function:} The metallic floating contact will act as a hot reservoir only if it has a well-defined quasi-equilibrated electronic distribution characterized by temperature $T_M$. For the realization of such a quasi-equilibrium regime, the hot electrons must dwell in metallic floating contact sufficiently longer than the electron-electron interaction (thermalization) time scale. The dwell time can be calculated as; $t_{dwell} = \frac{D_{E} \Omega h}{N}$\cite{jezouin2013quantum,brouwer1997charge}, where $D_{E}$ is the electronic density of states per unit volume per unit energy, $\Omega$ is the volume of micron-size floating contact, $h$ is the Planck's constant and $N$ is the number of channels leaving the floating contact. In typical graphene devices used for thermal conductance measurement, the floating contact has a volume ~$\Omega \approx 0.5 \mu$m$^3$, mostly made of gold. Using the typical density of states for gold $D_{E} \simeq 1.14\times10^{47}J^{-1}m^{-3}$, estimated dwell time was found to be $t_{dwell} \simeq \frac{40\mu s}{N}$, which is much larger than the typical electron-electron interaction time of the order of 10 ns range for gold at a temperature down to few milli-kelvin\cite{PhysRevB.68.085413}. This firmly establishes the quasi-equilibrium hypothesis that the electron's energy distribution in the micron size ohmic contact is a hot Fermi distribution function characterized by a temperature $T_M$.

\textbf{High-quality ohmic contact between the metallic floating reservoir and graphene channel:} As mentioned, since the electron temperature determination relies on excess thermal noise measurement, other unwanted noise sources should be avoided.  One such factor is that the metallic floating reservoir should make good quality ohmic contact with graphene to minimise the current reflection, which otherwise will generate unwanted Shot noise. Thanks to the development of a high-quality ohmic one-dimensional edge contact technique for graphene, the minimal reflection coefficient was obtained, which was negligible for all devices used for the thermal conductance measurements in literature.\\

\section{Universality of quantized heat flow in graphene}
\subsection{Thermal conductance of Integer quantum Hall states}
\begin{figure}[t!]
 	\centering
 	\includegraphics[width=1.0\textwidth]{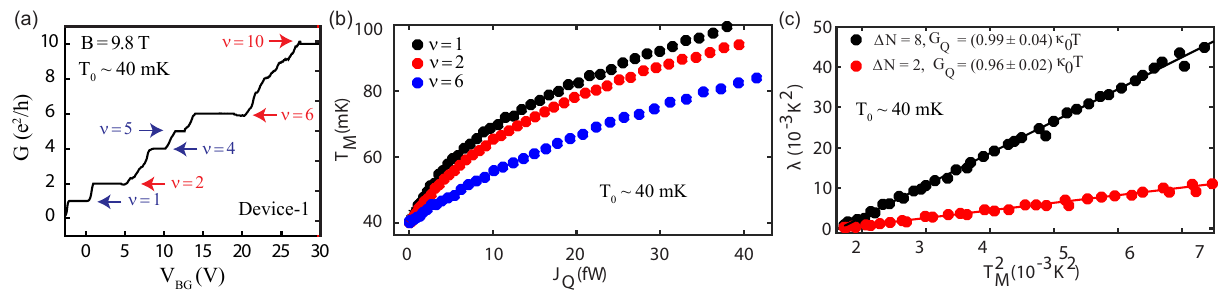}
 	\caption{ \textbf{(a)} Gate response of \ch{SiO_2/Si} gated hBN encapsulated graphene device at B = 9.8 T. Clear signature plateaus in conductance (marked with red arrows) at $\nu = 2, 6,$ and $10$ are observed in the unit of $\frac{e^2}{h}$. In addition to these, well-developed broken symmetric plateaus (marked with blue arrows) at $1\frac{e^2}{h}, 4\frac{e^2}{h},$ and $5\frac{e^2}{h}$ are also visible.  \textbf{(b)} The increased temperatures $T_{M}$ of the floating reservoir are plotted (solid circles) as a function of dissipated power $J_{Q}$ for $\nu$ = 1 ($N$ = 2), 2 ($n$ = 4) and 6 ($N$ = 12), respectively, where $n = 2\nu$ is the total outgoing channels from the floating reservoir. \textbf{(c)} The $\lambda = \Delta J_{Q}/(0.5 \kappa_{0})$ is plotted as a function of $T^2_{M}$ for $\Delta N = 2$ (between $\nu$ = 1 and 2), and 8 (between $\nu$ = 2 and 6), respectively in red and black solid circles, where $\Delta J_{Q} = J_{Q}(\nu_{i}, T_{M})-J_{Q}(\nu_{j}, T_{M})$. The solid lines are the linear fittings to extract the thermal conductance values. Slope of these linear fits are 1.92 and 7.92 for $\Delta N$ = 2, and 8, respectively, which gives the $G_{Q} = 0.96 \kappa _{0}T$, and $0.99 \kappa _{0}T$ for single edge mode, respectively.\textit{The figure is adapted from Srivastav et al. (2019) \cite{doi:10.1126/sciadv.aaw5798}. Reprinted with permission from AAAS}}
 	\label{fig:Srivastav_2019}
 \end{figure}

We will first discuss the results reported for the Integer quantum Hall states in graphene. Fig.~\ref{fig:Srivastav_2019}(a) shows the quantum Hall response of the device as a function of back-gate voltage at 9.8 T of the magnetic field at $\sim$ 40 mK of the bath temperature. In addition to the signature QH plateau of SLG at $2\frac{e^2}{h}, 6\frac{e^2}{h}, 10\frac{e^2}{h}$, QH plateaus at $1\frac{e^2}{h}, 4\frac{e^2}{h}, 5\frac{e^2}{h}$, are also visible, which are associated with the degeneracy lifting of zeroth and first Landau levels of the single layer graphene.

To measure the thermal conductance, the central floating contact was heated to a temperature $T_M$ due to the Joule dissipation $J_Q$ near the floating contact. The plot of $T_M$ against the $J_Q$ is shown in Fig.~\ref{fig:Srivastav_2019}(b). To extract the electronic contribution to the thermal conductance of the QH states, $J_Q$ can be plotted against the $T_M^2-T_0^2$ and the slope of the linear fit provides the value of the quantized thermal conductance $G_Q$ provided the electron-phonon cooling contribution $(J_Q^{e-ph})$ is negligible. Experimentally, It was found that $J_Q^{e-ph}$ is negligible for the typical device geometry till $\sim T_M=100$ mK\cite{doi:10.1126/sciadv.aaw5798}. However, to completely rule out any electron-phonon cooling ($J_Q^{e-ph}$) contribution, dissipated joule power between the two QH filling factors was subtracted at constant $T_M$ under the assumption that the $J_Q^{e-ph}$ does not depend on the number of edge channels leaving the floating contact. The resulting plot between $\lambda = \Delta J_Q/0.5\kappa_0$ (where $\Delta J_Q=J_Q(\nu_i, T_M)-J_Q(\nu_j, T_M)=0.5\Delta N\kappa_0 (T_M^2-T_0^2)$ ) vs $T_M^2-T_0^2$ is shown in Fig.~\ref{fig:Srivastav_2019}(c). From the slope of the curve $(m)$, thermal conductance per ballistic channel was calculated as $G_Q=(m/\Delta N)\kappa_0 T$ and found to be $\approx 1\kappa_0T$ as predicted theoretically. This plot is the first experimental demonstration of the quantized heat flow for integer quantum Hall states in graphene since its first mechanical exfoliation isolation. This experiment also emphasized that the universality of the quantized heat flow is the same for intrinsic QH states ($\nu=2, 6$) and symmetry-broken quantum Hall states ($\nu=1$) in graphene devices, as expected. 

\subsection{Thermal conductance of particle-like fractional quantum Hall states $(\nu=4/3)$}
The four-fold unique spin valley degeneracy of single-layer graphene distinguishes it from the conventional \ch{GaAs/AlGaAs} system. In the regime of fractional quantum Hall states, the presence of a large magnetic field can give rise to different possibilities of the symmetry breaking of the internal four-fold degeneracy, such as the complete lifting of the spin and valley degeneracy, hence the emergent FQH states will be either spin or valley polarised. Another possibility includes the degeneracy lifting of only one flavour, either spin or valley and hence, the emergent FQH phases preserve the approximate SU(2) symmetry in the remaining flavour sector. In addition to these two possibilities, there might be a scenario where none of the spin and valley flavours degeneracy is broken. Consequently, the emergent FQH phases have mixed spin-valley flavours. The earlier transport experiments and the local compressibility measurements suggest that while the FQH states with filling $0<\nu<1$ follow the standard sequence of non-interacting single component two flux composite fermion model, only the even numerator FQH states emerge for the filling $1<\nu<2$, suggesting the possibility of two-component composite fermion model, with an approximately SU(2) symmetric states for the interacting electrons. This complexity or richness of FQH states in graphene raised concern over the universality of the quantized heat flow for FQH states in graphene. 

\begin{figure}[t!]
 	\centering
 	\includegraphics[width=1.0\textwidth]{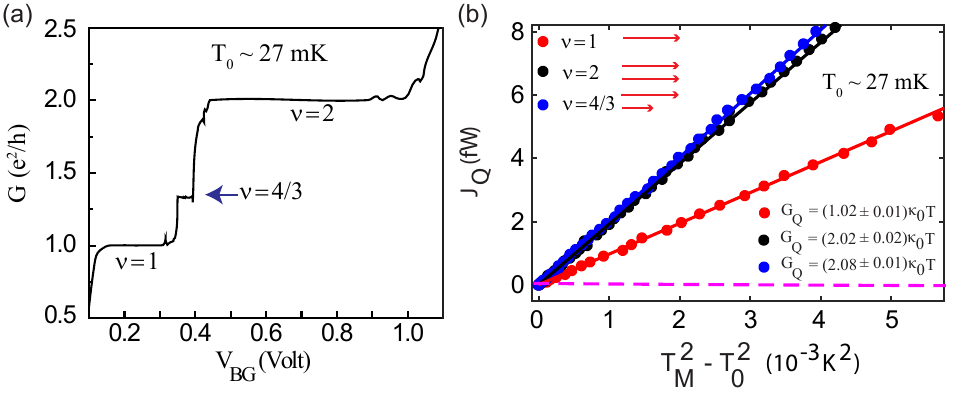}
 	\caption{ \textbf{(a)} Gate response of \ch{SiO_2/Si} gated hBN encapsulated graphene device at B = 9.8 T. Clear signature plateaus in conductance (marked with red arrows) at $\nu = 2, 6,$ and $10$ are observed in the unit of $\frac{e^2}{h}$. In addition to these, well-developed broken symmetric plateaus (marked with blue arrows) at $1\frac{e^2}{h}, 4\frac{e^2}{h},$ and $5\frac{e^2}{h}$ are also visible. \textbf{(b)} $J_Q$ (solid circles) is plotted as a function of $T^2_{M} - T^2_{0}$ for $\nu$ = 1, 4/3 and 2 and shown up to $T_M$ $\sim$ 60-70mK. The solid lines are the linear fit to extract the slopes, which give the thermal conductance values of 1.02, 2.08 and 2.02$\kappa _{0}T$ for $\nu$ = 1,  $\frac{4}{3}$, and 2, respectively. Thermal conductance values are quantized for $\nu$ = 1 and 2, and, more importantly, the values are the same for both $\nu$ = 4/3 and 2 plateaus. The inset shows the corresponding downstream charge modes for integer and fractional edges. The dashed curve represents the theoretically predicted contribution of heat coulomb blockade for $\nu$ = 1, showing its negligible contribution to the net thermal current. \textit{The figure is adapted from Srivastav et al. (2019) \cite{doi:10.1126/sciadv.aaw5798}. Reprinted with permission from AAAS}}
 	\label{fig:Srivastav_2019_FQH}
 \end{figure}

 The earlier experimental result of the heat flow measurement for the particle-like FQH state $(\nu=4/3)$\cite{doi:10.1126/sciadv.aaw5798} in an hBN encapsulated graphite back-gated single-layer graphene device is shown in Fig.~\ref{fig:Srivastav_2019_FQH}. Fig.~\ref{fig:Srivastav_2019_FQH}(a) shows the QH plateau of $\nu=4/3$ FQH state between the integer plateau of $\nu=1$ and $2$. Fig.~\ref{fig:Srivastav_2019_FQH}(b) shows the plot of $J_{Q}$ (solid circles) as a function of $T^2_{M} - T^2_{0}$ for $\nu$ = 1, $\frac{4}{3}$, and 2 over the temperature window where the curve is linear, implying the dominance of the electronic contribution to the heat flow. The solid lines in Fig.~\ref{fig:Srivastav_2019_FQH}(b) represent the linear fits (in $0.5\kappa_{0}$) and give the values of 2.04, 4.16 and 4.04, which corresponds to $G_{Q} = 1.02, 2.08$ and $2.02 \kappa _{0}T$ for $\nu$ = 1, $\frac{4}{3}$, and 2, respectively. Conventionally, $\nu$ = $\frac{4}{3}$, is thought of as $1+1/3$. As a result, one would expect two downstream charge modes, one integer and one fractional (inner $\nu$ = $\frac{1}{3}$ with effective charge, $e^* = \frac{e}{3}$). If the universality of quantized heat flow is preserved, the thermal conductance of $\nu =\frac{4}{3}$ should be the same as $\nu$ = 2 having two integer downstream charge modes. This is indeed the same as shown in Fig.~\ref{fig:Srivastav_2019_FQH}(b). For $\nu = 4/3$, $G_Q$ was found to be $(2.02\pm 0.02)\kappa_0T$. This measurement established the universality of the quantized thermal conductance in graphene for both integer and fractional QH edges.

\section{Non-equilibrated heat transport for hole-conjugate fractional quantum Hall states}
\begin{figure}[t!]
 	\centering
 	\includegraphics[width=1.0\textwidth]{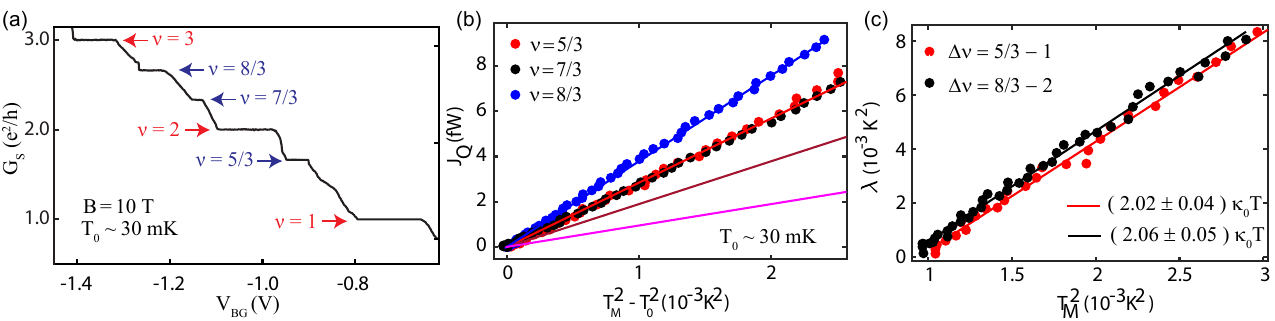}
 	\caption{\textbf{(a)} Conductance $G_S$ is plotted as a function of back gate voltage. The robust fractional plateaus at $\nu = \frac{5}{3}\frac{e^2}{h}$, $\frac{7}{3}\frac{e^2}{h}$, $\frac{8}{3}\frac{e^2}{h}$ with weaker plateau $\sim$ $\frac{4}{3}\frac{e^2}{h}$ clearly visible along with the integer QH plateaus at $\nu=1, 2,$ and 3. \textbf{(b)} $J_Q$ (solid circles) as a function of $T^2_{M} - T^2_{0}$ for $\nu$ = 5/3 (red), 7/3 (black) and 8/3 (blue). The solid magenta, brown, red and blue lines represent $G_{Q}$ = 1$\kappa _{0}T$, 2$\kappa _{0}T$, 3$\kappa _{0}T$ and 4$\kappa _{0}T$, respectively. The linear fits of the solid circles give $G_{Q}$ = 3.03, 2.96 and 4.03$\kappa _{0}T$ for $\nu$ = 5/3, 7/3 and 8/3, respectively. \textbf{(b)} $\lambda = \Delta J_{Q}/(0.5 \kappa_{0})$ as a function of $T^2_{M}$ for $\Delta \nu = 5/3-1$ (red) and $\Delta \nu = 8/3-2$ (black), where $\Delta J_{Q} = J_{Q}(\nu_{i},T_{M})-J_{Q}(\nu_{j},T_{M})$. Solid lines represent linear fits.  Extracted values of $G_{Q}$ of the $2/3$-like FQH states are $2.02\kappa _{0}T$ and $2.06\kappa _{0}T$ for $\Delta \nu = 5/3-1$ and $\Delta \nu = 8/3-2$, respectively. \textit{The figure is adapted from Srivastav et al. (2021) \cite{PhysRevLett.126.216803}. Reprinted with permission from American Physical Society}.}
 	\label{fig:Srivastav2021}
 \end{figure}

Till now, we have mostly discussed the thermal conductance for the integer and particle-like FQH states, where only downstream edge modes exist. However, the situation becomes complex for hole-conjugate FQH states, where the edge structure is complicated and hosts the downstream and upstream edge modes. One such paradigmatic FQH phase emerges at $\nu=2/3$, consisting of counter-propagating $1$ (downstream) and $\frac{1}{3}$ (upstream) modes~\cite{macdonald1990edge}. Although the thermal conductance for 2/3 state was measured in widely studied \ch{GaAs/AlGaAs} structure\cite{banerjee2017observed}, nothing was known in the case of graphene for hole-conjugate states. The first experimental report on the measurement of thermal conductance for hole-conjugate FQH states $(\nu=5/3, 8/3)$ in bilayer graphene was reported in 2021\cite{PhysRevLett.126.216803}, which shows a remarkably different result. Fig.\ref{fig:Srivastav2021}(a) shows the plot of the conductance versus gate voltage of an hBN encapsulated graphite gated bilayer graphene device at $B=10$ T and a bath temperature of 30 mK in the hole-doping region. In addition to the integer QH plateaus at $\nu=1,2, $ and $3$, well developed FQH plateau emerges at $\nu= 4/3, 5/3,$ and $8/3$. Conventionally, 5/3 and 8/3 states can be considered 1+2/3 and 2+2/3 states, respectively. 
 So, normally, one would expect that edge the structure of $\nu=5/3(8/3)$ should host one(two) downstream integer modes of conductance $e^2/h$, one downstream fractional edge mode of conductance $2/3(e^2/h)$ and one upstream neutral modes, which does not carry any charge but can support the heat flow. Counting of all these edge modes provides $N_d=2$ and $N_u=1$ for 5/3 state, and  $N_d=3$ and $N_u=1$ for 8/3 state, respectively. If one naively expects the full thermal equilibration between these edge modes, $G_Q$ for these hole-conjugate FQH states is expected to be $|N_{d}-N_{u}|\kappa _{0}T$, which should be $1\kappa_{0}T$ and $2\kappa_{0}T$, for 5/3 and 8/3, respectively. Fig.\ref{fig:Srivastav2021}(b) shows the plot of $J_Q$ with $T^2_{M} - T^2_{0}$ for $\nu$ = 5/3 (red), 7/3 (black) and 8/3 (blue). The filled circles are the experimental data points and the solid magenta, brown, red and blue lines represent the theoretical lines of $J_Q$ for $G_{Q}$ = 1$\kappa _{0}T$, 2$\kappa _{0}T$, 3$\kappa _{0}T$ and 4$\kappa _{0}T$, respectively. As clearly evident from the plots(See Fig.~\ref{fig:Srivastav2021}(b)), the measured $G_Q$  for $\nu=5/3,$ and $8/3$ strikingly matches with $(N_{d}+N_{u})\kappa _{0}T$ ($3\kappa_0T$ for 5/3 and $4\kappa_0T$ for 8/3), rather than the expected topological quantum number of $|N_{d}-N_{u}|\kappa _{0}T = 1\kappa _{0}T$, and $2\kappa _{0}T$, respectively. Fig.~\ref{fig:Srivastav2021}(c) shows the plot $\lambda = \Delta J_{Q}/(0.5 \kappa_{0})$ as a function of $T^2_{M}$ for two different configurations of $\Delta \nu = \frac{5}{3}-1$ (red) and $\frac{8}{3}-2$ (black) to extract the contribution of the partially filled Landau level with $\nu = $ $\frac{2}{3}$ out of the data for $\frac{5}{3}$ and $\frac{8}{3}$. Linear fits give $2.02\kappa _{0}T$ and $2.06\kappa _{0}T$, respectively, for $G_{Q}$ of the $\nu = $ $\frac{2}{3}$ state. This value is markedly inconsistent with the values dictated by the topology (0$\kappa_0T$ in long length limit) and the observation of Banerjee et al. (2017) in GaAs/ AlGaAs sample, which report the $G_Q\approx 0.33 \kappa_0T$ and interpret the equilibrated heat flow of the counter-propagating edge modes expected for $2/3$ state, i.e. $G_{Q} = (N_d - N_u)\kappa_{0}T$. It is important to emphasize that the measured value of the electrical conductance for 5/3 and 8/3 states in bilayer graphene in this experiment matches very well with the expected value of the equilibrated value of the electrical conductance. The theoretical study of this observed phenomenon in bilayer graphene pointed towards the diverging thermal equilibration length $l_{eq}^{H}$ while charge equilibration length $l_{eq}^{C}$ remains finite near the disorder fixed point\cite{PhysRevLett.126.216803}. The same experiment was repeated for the fractional states observed in the electron doping regime, and the observed results were the same. This is particularly important because the fractional states at the same fillings observed in electron and hole doping correspond to a different set of the orbital index in bilayer graphene. This experiment provides the first experimental notion of the different charges and the heat equilibration length for the counter-propagating edge modes in QH phases. A similar result was also reported by G. L. Bretona et al. \cite{le2022heat} in the first cool-down of their device.

 \section{Determination of topological edge quantum numbers}
\begin{figure}[t!]
 	\centering
 	\includegraphics[width=1.0\textwidth]{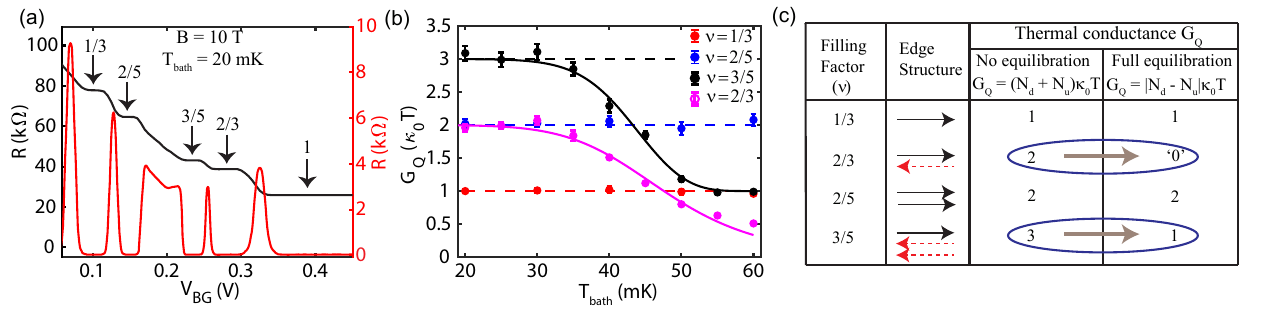}
 	\caption{\textbf{(a)} QH response of the device. Robust fractional plateaus at $\nu = \frac{1}{3}\frac{e^2}{h}$, $\frac{2}{5}\frac{e^2}{h}$, $\frac{3}{5}\frac{e^2}{h}$, and $\sim$ $\frac{2}{3}\frac{e^2}{h}$ clearly visible along with the integer QH plateau at $\nu=1$ (black curve). The vanishing longitudinal resistance (red curve) accompanies the QH plateaus at these fillings.  \textbf{(b)}Thermal conductance $G_{Q}$, plotted as a function of the bath temperature for $\nu = $ 1/3 (red), 2/5 (blue), 3/5 (black), and 2/3 (magenta). The horizontal dashed lines correspond to quantized $G_{Q}$ values. The solid curves (black and magenta) are theoretical fits of the data to extract out temperature scaling exponents of thermal equilibration length. \textbf{(c)} Edge structures of the studied FQH states. Solid black and dashed red arrows represent downstream and upstream modes, respectively. The two right-most columns show expected values of the thermal conductance $G_Q$ (in units of $\kappa_0 T$) in the two limiting regimes of the heat transport. \textit{The figure is adapted from Srivastav et al. (2022) \cite{srivastav2022determination}. Reprinted with permission from Springer Nature}.}
 	\label{fig:Crossover_G_Q_with_T}
 \end{figure}

 Although the measured values of $G_Q=$ $3\kappa_0T$ and $4\kappa_0T$ for 5/3 and 8/3 states, respectively, in bilayer graphene by Srivastav et al. (2021) are well explained by the non-equilibrated values of the heat flow of counter-propagating edge modes for 2/3 edge structure, it does not entirely rule out the possibility of two co-propagating downstream edge modes of electrical conductance of 1/3 $e^2/h$ per edge mode\cite{beenakker1990edge}. Measuring the electrical or thermal conductance quantization is essential to determine its edge structure. For example, non-equilibrated and equilibrated charge transport for 2/3 FQH state with counter-propagating edge modes should correspond to the two terminal electrical conductances of 4/3 $e^2/h$ and 2/3 $e^2/h$, respectively. However, experimentally, such a crossover of electrical conductance between two asymptotic limits was never observed for intrinsic 2/3 bulk filling. This may correspond to either the charge equilibration length being very short in conventional devices and limiting such crossover or the 2/3 edge structure comprising two downstream edge modes of electrical conductance 1/3 $e^2/h$ each. This dilemma was answered in the temperature-dependent study of quantized thermal conductance for 1/3, 2/5, 2/3, and 3/5 FQH states in a single-layer graphene device\cite{srivastav2022determination}, which is discussed now. Fig.~\ref{fig:Crossover_G_Q_with_T}(a) shows the gate response of the extremely clean hBN encapsulated graphite gated single-layer graphene device at B=10 T. Robust fractional plateaus at $\nu = \frac{1}{3}\frac{e^2}{h}$, $\frac{2}{5}\frac{e^2}{h}$, $\frac{3}{5}\frac{e^2}{h}$, and $\sim$ $\frac{2}{3}\frac{e^2}{h}$ clearly visible along with the integer QH plateau at $\nu=1$ (black curve). These QH plateaus are accompanied by the vanishing longitudinal resistance (red curve), establishing the robustness of the FQH states. The thermal conductance for the fractional states for $\nu=$ $1/3$ (red), $2/5$ (blue), $2/3$ (magenta), and $3/5$ (black) is plotted as a function of the bath temperature in Fig.~\ref{fig:Crossover_G_Q_with_T}(b). As evident from this plot, for $\nu=$ $1/3$ (red) and $2/5$ (blue) (no counter-propagating edge modes) the values $G_{Q}$ ($1\kappa_{0}T$ and $2\kappa_{0}T$, respectively) remain independent of the bath temperature.  On the other hand, hole-conjugate states showed surprising results. $G_Q$ for 2/3 and 3/5 (hole-like states with CP modes) was found to be $2\kappa_0T$ and $3\kappa_0T$ respectively, at 20 mK of bath temperature and match well with non-equilibrated regimes $(N_d + N_u)\kappa_{0}T$. Surprisingly, as the temperature increases, $G_Q$ starts decreasing, eventually taking the value of the equilibrated regime $(N_d - N_u)\kappa_{0}T$. In contrast to these hole-like fillings, $G_Q$ at 1/3 and 2/5 fillings (without CP modes) remains robustly quantized at $N_d \kappa_{0}T$ independent of temperature. 

 These results can be understood from the expected edge structures and their corresponding thermal conductance values for the studied FQH states in Fig.~\ref{fig:Crossover_G_Q_with_T}(c). For the electron-like $1/3$ and $2/5$ states, there are only downstream modes with $N_d = 1$ and $2$, respectively, and thus, the expected $G_Q$ should be $1\kappa_{0}T$ and $2\kappa_{0}T$, respectively, and should remain independent of the temperature. This is seen in the experiment as evident from Fig.~\ref{fig:Crossover_G_Q_with_T}(b). This behaviour is analogous to integer QH states, where all edge modes propagate downstream.
In contrast, for the hole-like $3/5$ state, the temperature dependence crossover of $G_Q$ from one quantum value to another rules out the possibility of having only downstream modes. Furthermore, the measured values of $3\kappa_{0}T$ and $1\kappa_{0}T$, respectively, perfectly match with the non-equilibrated $((N_d + N_u)\kappa_0 T)$ and equilibrated $(|N_{d}-N_{u}|\kappa_0 T)$ regimes of $G_Q$ with $N_d = 1$ and $N_u = 2$. Similarly, for $2/3$, experimental observation rules out the theoretical model with only downstream modes and supports the crossover from the non-equilibrated regime of $G_{Q}$ to the equilibrated regime with $N_d=N_u = 1$. The equilibrated transport in this situation is diffusive, so $G_Q$ is expected to tend to zero relatively slowly (as $\sim 1/L$) in the long-length limit. Since the device channel length $L$ used in experiment\cite{srivastav2022determination} is limited to $\sim 5$ $\mu m$, it is presumably the reason for a finite value of $\sim 0.5 \kappa_{0}T$ at $T_{bath} \sim 60$ mK. Thus, measuring the quantized values of $G_Q$ in two regimes determines the exact edge quantum number and, hence, the topological order of the FQH states. Such a systematic crossover study of $G_Q$ between its asymptotic regime of heat equilibration opens a new route for finding the topological order of exotic even-denominator FQH states, like 5/2, 3/2, and so on. Recently, G. L. Bretona et al. \cite{le2022heat} also reported a transition from a non-equilibrated heat regime to an equilibrated heat regime of $G_Q$ for 8/3 states in a graphene device. This crossover was achieved in a different cooldown of the same sample.

\subsection{Absence of edge reconstruction}
In quantum Hall physics, the issue of edge reconstruction may raise concerns over the validity of the bulk-edge correspondence principle. Specifically, it has been proposed that edge reconstruction might occur in various QH states\cite{Wan2002reconstruction,Wan2003edge}. Here, it should be emphasized that the experimental crossover of thermal conductance for hole-conjugate FQH states in two regimes and the temperature independence of thermal conductance for particle-like FQH states also suggest the absence of edge reconstruction in graphene devices. For example, for the $\nu=1/3$ state, the edge reconstruction would increase the total number of modes from 1 to 3. This would lead to a crossover from $G_Q=3\kappa_{0}T$ at lower temperatures (non-equilibrated regime) to $G_Q=1\kappa_{0}T$ at higher temperatures (equilibrated regime), which is very different from what was observed ($G_Q=1\kappa_{0}T$ at all temperatures). Similarly, for the $\nu=2/3$ state, edge reconstruction would increase the number of modes from 2 to 4. This would mean that, at low temperatures (the non-equilibrated regime), the thermal conductance value will be $G_Q=4\kappa_{0}T$. By the same token, for the $\nu=3/5$ edge (proposed to have $3$ edge channels), non-equilibrated heat transport would give rise to $G_Q=5\kappa_{0}T$. No traces of these values were observed in our measurements. These results strongly suggest the absence of edge reconstruction in the measured graphene devices.

\section{Conclusions}
The quantized thermal conductance measurement is a very powerful technique for determining the topological order of the complex fractional QH phases. The experimental verification of the universality of quantized thermal conductance in graphene is an exciting development. The results covered in this review are a remarkable manifestation of an interplay of equilibration and topology in FQH transport. The charge transport in QH in graphene is always in the equilibrated regime in the existing devices to date, the heat transport shows a crossover between the non-equilibrated to the equilibrated heat transport regime. Both asymptotic limits of the thermal equilibration are encoded by topologically quantized heat conductances determined by the topological edge quantum numbers. The experimental reports discussed in this review should also be relevant to other FQH states realized in different host materials. Particularly, the temperature dependence crossover of quantized thermal conductance in two asymptotic regimes of the thermal equilibration can be used to settle the debate on the ground state of non-Abelian $\nu=5/2$ FQH states. Till now, the interpretation of the measured thermal conductance  $\frac{5}{2}\kappa_0 T$ at $\nu=5/2$ is based on the assumptions about the presence, absence, or partial character of thermal equilibration~\cite{PhysRevB.99.085309,Simon2020,Asasi2020,park2020noise}.  In particular, the technique discussed in this review can also probe the thermal conductance of various quantum anomalous Hall phases observed recently in twisted bilayer graphene and twisted \ch{MoTe_2} systems. 

\section{Acknowledgement}
S. K. S. acknowledges the financial support from the Prime Minister Research Fellowship (PMRF) by the Ministry of Education, Govt. of India during the PhD tenure. S. K. S. further acknowledges the financial support from the Israel Academy of Sciences and Humanities (IASH) and Council for Higher Education (CHE) Excellence Fellowship. S. K. S. also thanks the Feinberg Graduate School, Weizmann Institute of Science for the Dean of Faculty Fellowship and Weizmann Postdoctoral Excellence Fellowship. A.D. thanks the Department of Science and Technology (DST) and Science and Engineering Research Board (SERB), India, for financial support (SP/SERB-22-0387) and acknowledges the Swarnajayanti Fellowship of the DST/SJF/PSA-03/2018-19. A.D. also thanks CEFIPRA project SP/IFCP-22-0005.


\end{document}